\documentclass[floatfix,twocolumn,aps,showpacs,showkeys,nofootinbib]{revtex4}

\usepackage{amssymb}
\usepackage{amsmath}
\usepackage{graphics}
\usepackage{epsfig}
\usepackage[dvips]{color}
\usepackage{bm}
\usepackage{color}

\newcommand{\be}{\begin{equation}}

\newcommand{\ee}{\end{equation}}
\newcommand{\bea}{\begin{eqnarray}}
\newcommand{\eea}{\end{eqnarray}}
\newcommand{\beas}{\begin{eqnarray*}}
\newcommand{\eeas}{\end{eqnarray*}}

\newcommand{\nn}{\nonumber\\}
\newcommand{\slsh}[1]{{\not \! #1}}
\newcommand{\slshh}[1]{{\not \!\! #1}}

\begin{document}
\title{The magnetized effective QCD phase diagram}
\author{Alejandro Ayala$^{1,2}$, C. A. Dominguez$^2$, L. A. Hern\'andez$^2$, M. Loewe$^{3,4,2}$, R. Zamora$^3$}
\affiliation{$^1$Instituto de Ciencias
  Nucleares, Universidad Nacional Aut\'onoma de M\'exico, Apartado
  Postal 70-543, M\'exico Distrito Federal 04510,
  Mexico.\\
  $^2$Centre for Theoretical and Mathematical Physics, and Department of Physics,
  University of Cape Town, Rondebosch 7700, South Africa\\
  $^3$Instituto de F\1sica, Pontificia Universidad Cat\'olica de Chile,
  Casilla 306, Santiago 22, Chile.\\
  $^4$Centro Cient\1fico-Tecnol\'ogico de Valpara'so, Casilla 110-V, Valpara\1so, Chile.}

\begin{abstract}
The QCD phase diagram in the temperature versus quark chemical potential plane is studied in the presence of a magnetic field, using the linear sigma model coupled to quarks. It is shown that the decrease of the couplings with increasing field strength obtained in this model  leads to the critical temperature for the phase transition to decrease with increasing field intensity (inverse magnetic catalysis). This happens provided that plasma screening is properly accounted for. It is also found that with increasing field strength the location of the critical end point (CEP) in the phase diagram moves toward lower values of the critical quark chemical potential and larger values of the critical temperature. In addition, the CEP approaches the temperature axis for  large values of the magnetic field. We argue that a similar behavior is to be expected in QCD, since the physical impact of the magnetic field, regardless of strength, is to produce a spatial dimension reduction, whereby virtual quark-antiquark pairs are closer on average and thus, the strength of their interaction decreases due to asymptotic freedom.

\end{abstract}



\maketitle

\section{Introduction}
\label{Introduction}
The behavior of strongly interacting matter in presence of magnetic fields has become a subject of increasing interest over the past few years. This interest has been sparked by recent lattice QCD (LQCD) results showing that the transition temperature with 2 + 1 quark flavors decreases with increasing magnetic field and that the strength of the condensate decreases for temperatures close and above the phase transition~\cite{Fodor,Bali:2012zg,Bali2,Bali3}. This behavior has been dubbed {\it inverse magnetic catalysis}, and it has been the center of attention in a large number of model-dependent analyses~\cite{Noronha,Bruckmann:2013oba,Ferreira,Fukushima,Mueller,Andersen2,Braun,Bruckmann,Ferrer,Fayazbakhsh,Farias, Ferreira1}. For recent reviews see Refs.~\cite{Andersenreview, Miransky}. In general terms, it seems that inverse magnetic catalysis is not obtained in mean field approaches describing the thermal environment~\cite{Andersen, Fraga, Loewe, Agasian, Mizher, Fraga2}, nor when calculations beyond mean field do not include magnetic effects on the coupling constants~\cite{ahmrv}.

The novel feature implemented in effective models, able to account self-consistently for inverse magnetic catalysis, is the decrease of the coupling constants with increasing field strength obtained from the model itself~\cite{amlz, Ayala2} without resorting to {\it ad hoc} parametrizations. This has been achieved within the Abelian Higgs model and the linear sigma model coupled to quarks (LSMq). This behavior is made possible by accounting for the screening properties of the plasma, which have been recently formulated consistently for theories with spontaneous symmetry breaking~\cite{ahmrv}. This results in a formalism beyond the mean field approximation~\cite{amlz}. Screening is also important to obtain a decrease of the coupling constant with the magnetic field strength in QCD in the Hard Thermal Loop approximation~\cite{Ayala3}. It has also been shown, by means of a QCD sum rules analysis at zero temperature, that both, the threshold energy for the onset of the continuum in the quark vector current spectral density (a phenomenological parameter that signals the onset of deconfinement) and the gluon condensate increase with increasing magnetic field, as expected~\cite{adhlrv}.

Recently, the LSMq has been also used to explore the phase diagram without magnetic fields~\cite{mich}. It was found that there are values for the model couplings that allow locating a critical end point (CEP) in the region where lattice inspired calculations find it~\cite{sharma}. Since the LSMq does not exhibit confinement, this behavior is attributed to the proper treatment of plasma screening, instead of to the existence of a given confinement length scale~\cite{roberts}.

A pertinent question is whether the above description in the presence of a magnetic field can be used to study how such CEP changes with the field intensity.  Recent LQCD calculations~\cite{Endrodimag} show that for very strong magnetic fields, inverse magnetic catalysis prevails and the phase transition becomes first order at asymptotically large values of the magnetic field for vanishing quark chemical potential $\mu$. A similar behavior is obtained in the Nambu Jona-Lasinio model if one includes a magnetic field dependence of the critical temperature in agreement with LQCD~\cite{Costa}.

In this paper we use the LSMq to explore the consequences of a proper handling of the plasma screening properties in the description of the magnetized effective QCD phase diagram. We show that when including self-consistently magnetic field effects in the calculation of both the effective potential as well as on the thermo-magnetic dependence of the coupling constants, the CEP's location moves toward smaller values of the critical quark chemical potential, and larger values of the critical temperature. In addition, above a certain value of the field strength the CEP moves to towards the $T$-axis. We argue that this behavior can be understood on general grounds, as the magnetic field produces a dimension reduction, whereby virtual charged particles from the vacuum are effectively constrained to occupy Landau levels, thus restricting their motion to a plane. This makes these particles to lay closer to each other on average, thus reducing the interaction strength for strongly coupled theories. This situation takes place regardless of how weak the external field is. 

The paper is organized as follows: In Sec.~\ref{efectivepotential} we recall the basic features of the linear sigma model with quarks, and in the presence of a magnetic field. Next, we write the effective potential at finite temperature and quark chemical potential, including all the degrees of freedom of the model. In Sec.~\ref{III} we find the thermo-magnetic corrections to the boson self-coupling and to the coupling between fermions and bosons. In Sec.~\ref{IV} we find the effects of the thermo-magnetic dependence of the couplings on the critical temperature $T_c$ for chiral symmetry restoration transition. We find that $T_c$ is a decreasing function of the magnetic field for any value of $\mu$. We also study the magnetized effective QCD phase diagram and in particular find how the CEP's location changes with the field strength. We finally summarize and conclude in Sec.~\ref{conclusions}. In the appendix we describe the computation of the thermo-magnetic corrections to the couplings, and to the temperature and density dependence of the fermion thermal mass.

\section{Effective potential}\label{efectivepotential}

The Lagrangian of the sigma model, including quark degrees of freedom, is given by 
\begin{eqnarray}
   \mathcal{L}&=&\frac{1}{2}(\partial_\mu \sigma)^2  + \frac{1}{2}(D_\mu \vec{\pi})^2 + \frac{a^2}{2} (\sigma^2 + \vec{\pi}^2) - \frac{\lambda}{4} (\sigma^2 + \vec{\pi}^2)^2 \nonumber \\ 
   &+& i \bar{\psi} \gamma^\mu D_\mu \psi -g\bar{\psi} (\sigma + i \gamma_5 \vec{\tau} \cdot \vec{\pi} )\psi ,
\label{lagrangian}
\end{eqnarray}
where $\psi$ is an SU(2) isospin doublet, $\vec{\pi}=(\pi_1, \pi_2, \pi_3 )$ is an isospin triplet and $\sigma$ is an isospin singlet, with
\be
   D_{\mu}=\partial_{\mu}+iqA_{\mu},
\label{dcovariant}
\ee
is the covariant derivative. $A^\mu$ is the vector potential corresponding to an external magnetic field directed along the $\hat{z}$ axis. In the symmetric gauge it is given by
\be
   A^\mu=\frac{B}{2}(0,-y,x,0),
\label{vecpot}
\ee
where $q$ is the particle's electric charge. $A^\mu$ satisfies the gauge condition $\partial_\mu A^\mu=0$. The gauge field couples only to the charged pion combinations, namely
\be
   \pi_\pm=\frac{1}{\sqrt{2}}\left(\pi_1\mp i\pi_2\right).
\ee
The neutral pion is taken as the third component of the pion isovector, $\pi^0=\pi_3$. The gauge field is taken as classical and thus we do not consider loops involving the propagator of the gauge field in internal lines. The squared mass parameter $a^2$ and the self-coupling $\lambda$ and $g$ are taken to be positive.

To allow for spontaneous symmetry breaking, we let the $\sigma$ field to develop a vacuum expectation value $v$
\be
   \sigma \rightarrow \sigma + v,
\label{shift}
\ee
This vacuum expectation value can later be identified as the order parameter of the theory. After this shift, the Lagrangian can be rewritten as
\bea
   {\mathcal{L}} &=& -\frac{1}{2}[\sigma(\partial_{\mu}+iqA_{\mu})^{2}\sigma]-\frac{1}
   {2}\left(3\lambda v^{2}-a^{2} \right)\sigma^{2}\nn
   &-&\frac{1}{2}[\vec{\pi}(\partial_{\mu}+iqA_{\mu})^{2}\vec{\pi}]-\frac{1}{2}\left(\lambda v^{2}- a^2 \right)\vec{\pi}^{2}+\frac{a^{2}}{2}v^{2}\nn
  &-&\frac{\lambda}{4}v^{4} + i \bar{\psi} \gamma^\mu D_\mu \psi 
  -gv \bar{\psi}\psi + {\mathcal{L}}_{I}^b + {\mathcal{L}}_{I}^f,
  \label{lagranreal}
\eea
where ${\mathcal{L}}_{I}^b$ and  ${\mathcal{L}}_{I}^f$ are given by
\begin{eqnarray}
  {\mathcal{L}}_{I}^b&=&-\frac{\lambda}{4}\Big[(\sigma^2 + (\pi^0)^2)^2\nn 
  &+& 4\pi^+\pi^-(\sigma^2 + (\pi^0)^2 + \pi^+\pi^-)\Big],\nn
  {\mathcal{L}}_{I}^f&=&-g\bar{\psi} (\sigma + i \gamma_5 \vec{\tau} \cdot \vec{\pi} )\psi.
  \label{lagranint}
\end{eqnarray}
The terms given in Eq.~(\ref{lagranint}) describe the interactions among the fields $\sigma$, $\vec{\pi}$ and $\psi$, after symmetry breaking. From Eq.~(\ref{lagranreal}) one can see that the $\sigma$, the three pions and the quarks have masses given, respectively, by
\bea
  m^{2}_{\sigma}&=&3  \lambda v^{2}-a^{2},\nn
  m^{2}_{\pi}&=&\lambda v^{2}-a^{2}, \nn
  m_{f}&=& gv.
\label{masses}
\eea

Using Schwinger's proper-time method, the expression for the one-loop effective potential for one boson field with squared mass $m_b^2$ and absolute value of its charge $q_b$ at finite temperature $T$ in the presence of a constant magnetic field can be written as
\bea
  V_b^{(1)} &=& \frac{T}{2}\sum_n\int dm_b^2\int\frac{d^3k}{(2\pi)^3}\int_0^\infty
   \frac{ds}{\cosh (q_bBs)}\nn
   &\times&e^{-s(\omega_n^2+k_3^2 + k_\perp^2\frac{\tanh (q_bBs)}{q_bBs} + m_b^2)},
   \label{boson1}
\eea
where $\omega_n=2n\pi T$ are boson Matsubara frequencies.
Similarly, the expression for the one-loop effective potential for one fermion field with mass $m_f$ and absolute value of its charge $q_f$ at finite temperature $T$ and chemical potential $\mu$, in the presence of a constant magnetic field can be written as
\bea
  V_f^{(1)}&=& -\sum_{r=\pm 1}T\sum_n\int dm_f^2\int\frac{d^3k}{(2\pi)^3}\int_0^\infty
   \frac{ds}{\cosh (q_fBs)}\nn
   &\times&e^{-s[(\tilde{\omega}_n-i\mu)^2+k_3^2 + k_\perp^2\frac{\tanh (q_fBs)}{q_fBs} + m_f^2 +r q_fB]},
   \label{fermion1}
\eea
where $\tilde{\omega}_n=(2n+1)\pi T$ are fermion Matsubara frequencies. The sum over the index $r$ corresponds to the two possible spin orientations along the magnetic field direction. 

Including the $v$-independent terms, choosing the renormalization scale as $\tilde{\mu}=e^{-1/2}a$ and after mass and charge renormalization, the thermo-magnetic effective potential in the small to intermediate field regime, in a high temperature expansion can be written as
\bea
   V^{({\mbox{\small{eff}}})}&=&
   -\frac{a^2}{2}v^2 + \frac{\lambda}{4}v^4\nn
   &+&\sum_{i=\sigma,\pi^0}\left\{\frac{m_i^4}{64\pi^2}\left[ \ln\left(\frac{(4\pi T)^2}{2a^2}\right) 
   -2\gamma_E +1\right]
   \right.\nn
   &-&\left. \frac{\pi^2T^4}{90} + \frac{m_i^2T^2}{24}  - \frac{T}{12 \pi}(m_i^2 + \Pi)^{3/2} \right\} \nn
   &+&\sum_{i=\pi_+,\pi_-}\left\{\frac{m_i^4}{64\pi^2}\left[ \ln\left(\frac{(4\pi T)^2}{2a^2}\right) 
   -2\gamma_E +1\right]
   \right.\nn
   &-&\frac{\pi^2T^4}{90} + \frac{m_i^2T^2}{24}\nn
   &+&\frac{T(2qB)^{3/2}}{8\pi}\zeta\left(-\frac{1}{2},\frac{1}{2}+\frac{m_i^2+\Pi}{2qB}\right)\nn
   &-&\frac{(qB)^2}{192\pi^2}\left[ \ln\left(\frac{(4\pi T)^2}{2a^2}\right)
   - 2\gamma_E + 1\right.\nn
   &+&\left.\left. 
   \zeta (3)\left(\frac{m_i}{2\pi T}\right)^2 - \frac{3}{4}\zeta (5) \left(\frac{m_i}{2\pi T}\right)^4   
   \right]\right\} \nn 
   &-&N_c \sum_{f=u,d}\biggl[\frac{m_f^4}{16\pi^2} \biggl[ \ln\left(\frac{(4 \pi T)^2}{2a^2}\right) \nn
   &+& \psi^{0}\left(\frac{1}{2}+\frac{i \mu}{2\pi T}\right)+\psi^{0}\left(\frac{1}{2}-\frac{i \mu}{2\pi T}\right) \biggr] \nn
   &+& 8 m_f^2 T^2[Li_2(-e^{\mu/T})+Li_2(-e^{-\mu/T})] \nn   
   &-& 32T^4 [Li_4(-e^{\mu/T})+Li_4(-e^{-\mu/T})] \nn
   &+& \frac{(q_fB)^2}{24\pi^2} \biggl[ \ln\left(\frac{(\pi T)^2}{2a^2}\right) - 2\gamma_E + 1 \nn
   &-&\psi^{0}\left(\frac{1}{2}+\frac{i \mu}{2\pi T}\right)-\psi^{0}\left(\frac{1}{2}-\frac{i \mu}{2\pi T}\right) \nn
   &+& \frac{2\pi}{((\pi+i\mu/T)^2+m_f^2/T^2)^{1/2}} \nn
   &+&\frac{2\pi}{((\pi-i\mu/T)^2+m_f^2/T^2)^{1/2}}  \nn
   &-& \frac{4\pi}{(\pi^2+m_f^2/T^2)^{1/2}} \biggr] \biggr],
   \label{Veff-mid}
\eea
where $\psi^0(x)$ is the digamma function, $Li_n$ is the polylogarithm function of order $n$, $q$ is the absolute value of the charged pion charge ($q=1$), $q_u=2/3$, and $q_d=1/3$ are the absolute values of the \textit{up-} and \textit{down-}quarks, respectively, $\gamma_E$ is the Euler's gamma, $N_f=2$ is the number of light-quark flavors, and $N_c=3$ is the number of colors. In order to obtain the leading magnetic field contribution of Eq.~(\ref{Veff-mid}) we use the Euler-MacLaurin expansion~\cite{ahmrv} in Eqs.~(\ref{boson1}) and~(\ref{fermion1}). 

Though we take the quark masses as equal, the notation emphasizes that the effective potential is evaluated after accounting for the different quark charges. We have introduced the leading temperature plasma screening effects for the boson masses squared, encoded in the boson self-energy $\Pi$. The leading contribution for the boson self-energy in a high temperature expansion, and at finite $\mu$ is given by~\cite{mich}
\bea
   \Pi= \lambda \frac{T^2}{2} - N_f N_cg^2\frac{T^2}{\pi^2}[Li_2(-e^{\mu/T})+Li_2(-e^{-\mu/T})].\nn
\label{self}
\eea
For the Hurwitz zeta function $\zeta(-1/2,z)$ in Eq.~(\ref{Veff-mid}) to be real, one needs 
\bea
   -a^2 + \Pi > qB,
\label{requirealso}
\eea
a condition that arises from requiring that the second argument of the Hurwitz zeta function satisfies $z>0$, even for the lowest value of $m_b^2$ which is obtained for $v=0$. Furthermore, for the large $T$ expansion to be valid, one also requires
\bea
   qB/T^2 <1.
\label{otherrequirement}
\eea
The conditions expressed in Eqs.~(\ref{requirealso}) and~(\ref{otherrequirement}) provide the limits of applicability of the high temperature expansion of the effective potential in Eq.~(\ref{Veff-mid}).

\section{Thermo-magnetic couplings}\label{III}

\begin{widetext}
\begin{figure*}[ht]
\begin{center}
\includegraphics[scale=1]{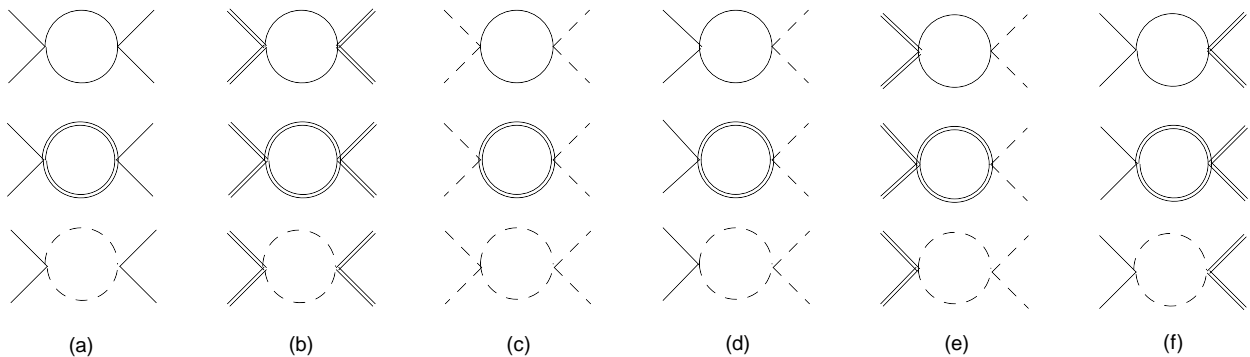}
\end{center}
\caption{One-loop Feynman diagrams that contribute to the thermal and magnetic correction to the coupling $\lambda$. The dashed line denotes the charged pion, the continuous line is the sigma and the double line represents
the neutral pion.}
\label{fig1}
\end{figure*}
\end{widetext}

We now compute the one-loop correction to the coupling $\lambda$, including thermal and magnetic effects. Figure~\ref{fig1} shows the Feynman diagrams that contribute to this correction. Columns (a), (b), (c), (d), (e) and (f) contribute, respectively, to the correction to the $\sigma^4$, $(\pi^0)^4$, $(\pi^+)^2(\pi^-)^2$, $\sigma^2\pi^+\pi^-$, $(\pi^0)^2\pi^+\pi^-$ and  $\sigma^2(\pi^0)^2$ terms of the interaction Lagrangian in Eq.~(\ref{lagranint}), respectively. Since each of these corrections lead to the same result, we concentrate on the diagrams in column (a). Each of the three diagrams involves two propagators of the same boson. For the first two diagrams the intermediate bosons are neutral and for the third one the intermediate bosons are charged. 
%%%%%%%%%%%%%%%%%%%%%%%%%%%%%%%%%%%%%%%%%%%%%
%%%%%%%%%%%%%%%%%%%%%%%%%%%%%%%%%%%%%%%%%%%%%
The explicit computation has been performed in Ref.~\cite{Ayala2}. This involves the use of the weak field expansion of the charged boson Schwinger proper-time propagator for the intermediate charged particles and its corresponding $qB\rightarrow 0$ limit for the intermediate neutral bosons. Since the analysis of Ref.~\cite{Ayala2} was carried out in the very high temperature case, the four-boson vertex correction was evaluated in the {\it infrared limit}, namely $P_i=(0,{\mbox{\bf{p}}}\rightarrow 0)$, where $P_i$ are the momenta of each of the four external legs. However, for the present analysis, where $T$ is close to $T_c$ a more appropriate treatment is to evaluate the vertex function at the typical energy of the external particles. This corresponds to the {\it static limit}, namely $P_i=(\Pi,{\mbox{\bf{p}}}= 0)$, where $\Pi$ is given by Eq.~(\ref{self}) and represents the purely thermal (and density) component of the boson mass. In the appendix, we explicitly reproduce such calculation, evaluating the vertex function in the static limit. Notice that $\lambda_{\mbox{\small{eff}}}$ depends on $v$ through the dependence on the boson masses. We further consider the approximation where $\lambda_{\mbox{\small{eff}}}$ is evaluated at $v=0$ since we are pursuing the effect on the critical temperature, which is the temperature where the curvature of the effective potential at $v=0$ vanishes.

\begin{figure}[bh]
\begin{center}
\includegraphics[scale=0.48]{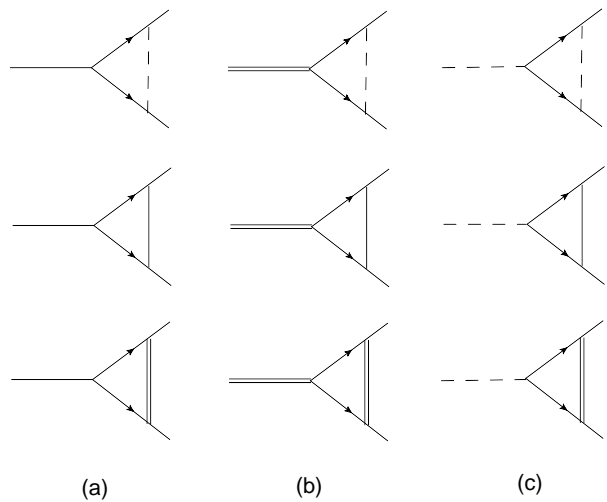}
\end{center}
\caption{One-loop Feynman diagrams that contribute to the thermal and magnetic correction to the coupling $g$. The dashed line denotes the charged pion, the continuous line is the sigma, the double line represents the neutral pion and the continuous line with arrows represents the quarks.}
\label{fig2}
\end{figure}

Next we turn to the calculation of the thermo-magnetic correction of the coupling $g$. Figure~\ref{fig2} shows the Feynman diagrams that contribute to this correction. We are interested in computing an effective value for this coupling, $g_{\mbox{\small{eff}}}$, as well as for $v = 0$, in the same manner done for $\lambda_{\mbox{\small{eff}}}$. Columns (a), (b) and (c) contribute to the correction to the quark-$\sigma$, quark-$\pi^0$ and quark-$\pi^\pm$ terms of the interaction Lagrangian of Eq.~(\ref{lagranint}), respectively. Since each of these corrections leads to the same result, we concentrate on the diagrams in column (a). Notice that for $v=0$ in Eq.~(\ref{masses}), the masses of the $\sigma$ and the $\pi^0$ become degenerate. Hence, the middle and bottom diagrams in column (a) of Fig.~\ref{fig2} cancel out, since they contribute with opposite signs. This also happens with the two bottom diagrams in columns (b) and (c). 

%%%%%%%%%%%%%%%%%%%%%%%%%%%%%%
%%%%%%%%%%%%%%%%%%%%%%%%%%%%%%
The explicit computation is carried out in the weak field limit of the charged boson and fermion Schwinger proper-time propagators. The calculation was done in Ref.~\cite{Ayala2} considering only terms of ${\mathcal{O}}$ $(qB)^2$ after taking the average over spins. We have repeated the calculation and realized that the contribution ${\mathcal{O}}$ $(qB)$ does not vanish. This is because such spin average is not needed, as we are not considering a decay process, but rather a vertex function. We have also evaluated $g_{\mbox{\small{eff}}}$ in the static limit $P_i=(m_f^{\mbox{\tiny{them}}},{\mbox{\bf{p}}}= 0)$ where $P_i$ are the momenta of the quark and antiquark and $m_f^{\mbox{\tiny{them}}}$ is the fermion thermal (and density)-dependent mass, which we compute in the appendix and whose square is given explicitly by
\bea
(m_f^{\mbox{\tiny{them}}})^2=g^2 T^2\left(\frac{1}{3}-\frac{Li_2(-e^{\mu/T})}{\pi^2}-\frac{Li_2(-e^{-\mu/T})}{\pi^2}\right).\nn
\eea

%%%%%%%%%%%%%%%%%%%%%%%%%%%%%%%%%%%%%%%%%%%%%%%%%%%%
\begin{figure}[th]
\begin{center}
\includegraphics[scale=0.6]{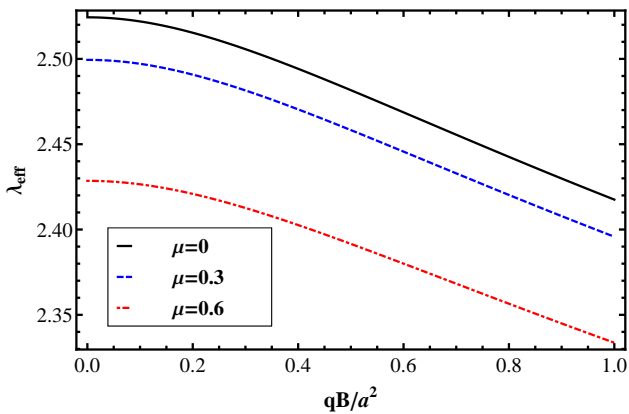}
\end{center}
\caption{Effective boson coupling $\lambda_{\mbox{\small{eff}}}$ evaluated a the temperature $T=180$ MeV with $\lambda=0.4$, $g=0.63$ as a function of the magnetic field strength for different values of $\mu$.}
\label{fig3}
\end{figure}
%%%%%%%%%%%%%%%%%%%%%%%%%%%%%%%%%%%%%%%%%%%%%%%%%%%%

To fix the bare values of the couplings $\lambda$, $g$ and $a$ appropriate for the description of the phase transition, we notice that the boson masses are modified when considering the thermal effects, since they acquire a thermal component. For $\mu=0$ they become
\bea
   m_\sigma^2(T)&=&3\lambda v^2 -a^2 + \frac{\lambda T^2}{2}+\frac{N_fN_cg^2T^2}{6}\nn
   m_\pi^2(T)&=&\lambda v^2 -a^2 + \frac{\lambda T^2}{2}+\frac{N_fN_cg^2T^2}{6}.\nonumber\\
\label{massmod} 
\eea 
At the phase transition, the curvature of the
effective potential vanishes for $v=0$. Since the boson 
masses are proportional to this curvature, these also vanish at
$v=0$. In this case, and from any of Eqs.~(\ref{massmod}), we then obtain a relation
between $T_c^0$ and the model parameters at the critical temperature with $\mu=0$
\bea
   \frac{a}{T_c^0}=\sqrt{\frac{\lambda}{2}+\frac{N_fN_cg^2}{6}}.
\label{relation}
\eea
Furthermore, we can fix the value of $a$ by noting from Eqs.~(\ref{masses}) that the vacuum boson masses satisfy
\bea
   a=\sqrt{\frac{m_\sigma^2 - 3m_\pi^2}{2}}.
\label{massvac}
\eea

%%%%%%%%%%%%%%%%%%%%%%%%%%%%%
\begin{figure}[th]
\begin{center}
\includegraphics[scale=0.6]{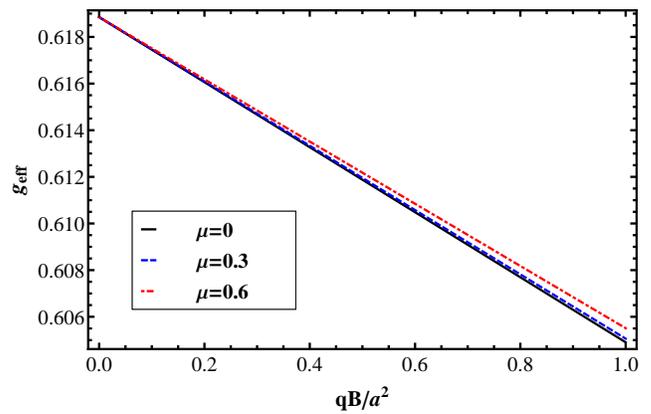}
\end{center}
\caption{Effective boson-fermion coupling $g_{\mbox{\small{eff}}}$ evaluated a the temperature $T=180$ MeV with $\lambda=0.4$, $g=0.63$ as a function of the magnetic field strength for different values of $\mu$.}
\label{fig4}
\end{figure}
%%%%%%%%%%%%%%%%%%%%%%%%%%%%%%
Since the effective potential is written as an expansion in powers of $a/T$ we need that this ratio satisfies $a/T<1$. From Eqs.~(\ref{relation}) and~(\ref{massvac}) the coupling constants are proportional to $m_\sigma$ which, from the above conditions, restricts the analysis to considering not too large values of $m_\sigma$ as well as not too small values of $T_c^0$. Since the purpose of this work is not to pursue a precise determination of the couplings but instead to call attention to the fact that the proper treatment of screening effects allows the linear sigma model to provide solutions for the CEP even at finite values of $\mu$, we consider small values for $m_\sigma$. Given that $\sigma$ is anyhow a broad resonance, in order to satisfy the above requirements let us  take for definitiveness $m_\sigma = 300$ MeV. namely, close to the two-pion threshold. For $T_c^{0}$ with two light quark flavors we take $T_c^{0}=172$ MeV~\cite{example}. Thus, $a/T_c^0=0.77$.
Equation~(\ref{relation}) provides a relation between $\lambda$ and $g$. A possible solution consistent with the above requirements is given by $\lambda=0.4$, $g=0.63$.

Figure~\ref{fig3} shows the behavior of the effective boson coupling $\lambda_{\mbox{\small{eff}}}$ evaluated using $T= 180$ MeV, as a function of magnetic field strength for different values of $\mu$. The considered temperature is slightly larger that $T_c^{0}$. Note that $\lambda_{\mbox{\small{eff}}}$ is a monotonically decreasing function of $qB$ and that the decrease is larger for larger values of $\mu$.

Figure~\ref{fig4} shows the behavior of $g_{\mbox{\small{eff}}}$ as a function of $qB$ evaluated also using $T=180$ MeV with $\lambda=0.4$, $g=0.63$ for three different values of $\mu$. Note that $g_{\mbox{\small{eff}}}$ is also a monotonically decreasing function of $qB$. However, the decrease is less pronounced for larger values of $\mu$. 
Note that the $\mu$-dependence of the effective coupling comes from its dependence on $m_f^{\mbox{\tiny{them}}}$.
 
\section{Inverse magnetic catalysis and the effective phase diagram}\label{IV}

%%%%%%%%%%%%%%%%%%%%%%%%%%%%%%
\begin{figure}[th]
\begin{center}
\includegraphics[scale=0.6]{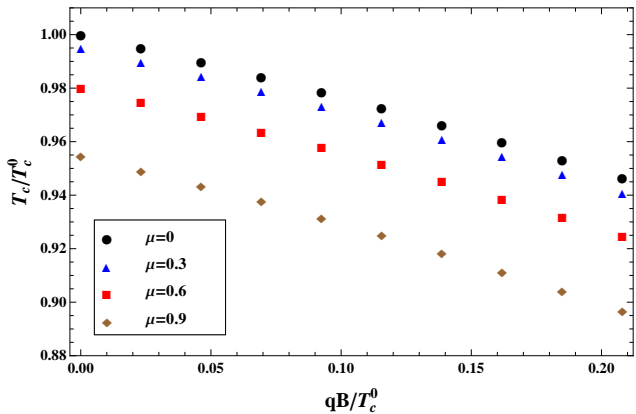}
\end{center}
\caption{Critical temperature as a function of the magnetic field strength evaluated using effective couplings including thermo-magnetic corrections with the bare values of the couplings $\lambda=0.4$, $g=0.63$ for different values of $\mu$. Note that in all cases the critical temperature is a monotonically decreasing function of the magnetic field strength.}
\label{fig5}
\end{figure}
%%%%%%%%%%%%%%%%%%%%%%%%%%%%%%
We now study the effect of the thermo-magnetic corrections to the couplings on the critical temperature.  For a given value of the magnetic field, and for a second order phase transition, the critical temperature is determined after setting to zero the second derivative of the effective potential at $v=0$ in Eq.~(\ref{Veff-mid}). When the phase transition becomes first order, the critical temperature is computed by determining the temperature where a secondary minimum for $v\neq 0$ is degenerate with a minimum at $v=0$. Figure~\ref{fig5} shows the critical temperature as a function of field strength, for different values of $\mu$, and for the bare values $\lambda=0.4$, $g=0.63$. Note that in all cases the critical temperature is a decreasing function of the field strength. 

\begin{figure}[th]
\begin{center}
\includegraphics[scale=0.6]{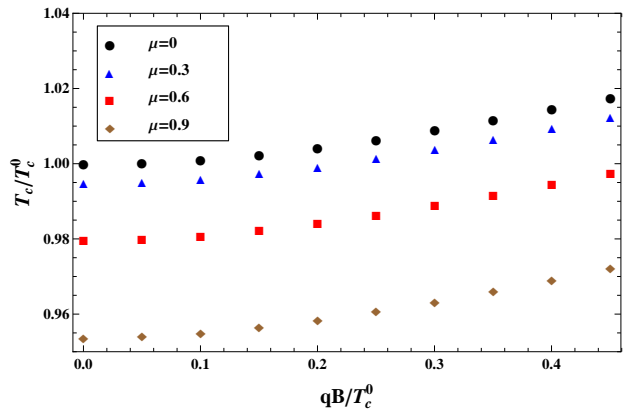}
\end{center}
\caption{Critical temperature as a function of the magnetic field strength evaluated without effective couplings and instead with the bare values of the couplings $\lambda=0.4$, $g=0.63$ for different values of $\mu$. Note that in all cases the critical temperature is a monotonically increasing function of the magnetic field strength.}
\label{fig6}
\end{figure}

On the contrary when the calculation is performed without including the thermo-magnetic modification to the couplings, the critical temperature turns out to be an increasing function of the field strength. This is shown in Fig.~\ref{fig6} where we plot the critical temperature as a function of $qB$ for different values of $\mu$ with the bare values of the couplings $\lambda=0.4$, $g=0.63$. This behavior shows that the thermo-magnetic corrections to the couplings are crucial to obtain inverse magnetic catalysis.

We now turn to describe the phase diagram in the temperature quark chemical potential plane. Figure~\ref{fig7} shows the phase diagram obtained for the bare couplings $\lambda=0.4$, $g=0.63$ for different values of the magnetic field strength. The thermo-magnetic corrections enter the analysis both in the effective potential of Eq.~(\ref{Veff-mid}), and in the effective couplings. Notice that as the field intensity increases, the CEP moves toward lower values of the critical quark chemical potential, and to larger values of the critical temperature and in this case, it reaches the $T$-axis. However, since our analysis is carried out in the weak field limit $qB/T^2 < 1$, we can only say that there is a tendency for the CEP to eventually reach the $T$-axis large values of the field strength.
%%%%%%%%%%%%%%%%%%%%%%%%%%%%%%
%%%%%%%%%%%%%%%%%%%%%%%%%%%%%%

To see the effect of a change of parameters we now explore the case where the ratio $a/T_c^0$ appearing in Eq.~(\ref{relation}) changes. We take $a/T_c^0=0.66$ which is obtained maintaining $m_\sigma=300$ MeV and increasing the value of $T_c^0$ to $T_c^0=200$ MeV. With this ratio, a possible solution to Eq.~(\ref{relation}) for the bare values of the couplings is given by $\lambda=0.36$, $g=0.51$. Figure~\ref{fig8} shows the phase diagram thus obtained. Notice that the CEP for $qB=0$ happens for values of $T_c^{\mbox{\tiny{CEP}}}$ and $\mu_c^{\mbox{\tiny{CEP}}}$ slightly smaller and larger, respectively, than for the corresponding values in Fig.~\ref{fig7}. When the magnetic field intensity increases the CEP also moves toward lower values of the critical quark chemical potential and larger values of the critical temperature but this time, for the largest value of $qB$ considered, the CEP does not reach the $T$-axis. Nevertheless we observe a tendency for the CEP to eventually reach the $T$-axis for larger values of $qB$ that can not be studied within the present small field approach.

\section{Summary and Conclusions}\label{conclusions}

In conclusion, we have shown that working in the LSMq in the presence of magnetic fields, it is possible to obtain values for the couplings that allow to locate a CEP that for $qB=0$, lays in the region found by mathematical extensions of lattice analyses. The analysis is done from the effective potential computed in the presence of a weak magnetic field and accounting for the plasma screening effects. Since the LSMq does not have confinement, we attribute the CEP's location to the adequate description of the plasma screening properties. Screening is included into the calculation in two manners: 
%%%%%%%%%%%%%%%%%%%%%%%%%%%%%%
\begin{figure}[th]
\begin{center}
\includegraphics[scale=0.6]{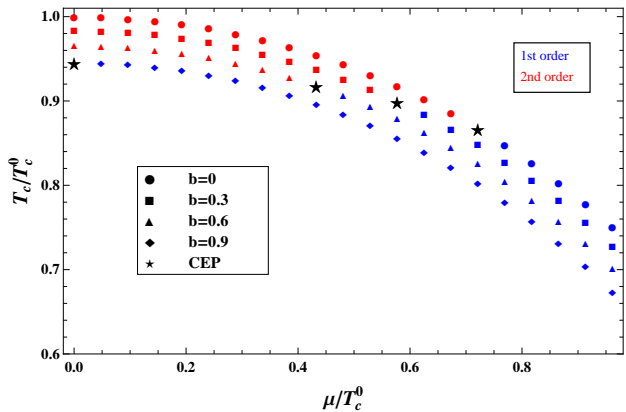}
\end{center}
\caption{Phase diagram in the temperature quark chemical potential plane computed with thermo-magnetic corrections to the couplings using the bare values $\lambda=0.4$, $g=0.63$ corresponding to $a/T_c^0=0.77$ for different values of the magnetic field strength. The phase transitions to the left (right) of the CEP in each case are of second (first) order.}
\label{fig7}
\end{figure}
%%%%%%%%%%%%%%%%%%%%%%%%%%%%%%
First, in the effective potential through the boson's self-energy and second in the thermo-magnetic corrections of the couplings. We have shown that this last correction is crucial to obtain inverse magnetic catalysis. To define the allowed range for the bare coupling constants we observe that the thermal boson masses vanish at the phase transition for $\mu=0$. This condition determines a relation between the model parameters which can be put in quantitative terms from knowledge of $T_c^0$ and $a$. The first can be obtained from lattice results and the second from the vacuum boson masses. Since the model is computed in the high temperature limit, we are limited to consider ratios of the parameter $a/T_c^0$ a bit off their usual values. Nevertheless, the model shows in quantative terms that the CEP moves toward lower values of the critical quark chemical potential and larger values of the critical temperature as the field intensity increases and that there is a tendency for the CEP to eventually reach the $T$-axis for a larger value of the field strength. 

The overall features of the phase diagram can be understood in general terms when we recall that the magnetic field produces a dimension reduction whereby the virtual charged particles that make up the vacuum are effectively constrained to occupy Landau levels which, in semiclassical terms, implies that their motion is restricted to planes. This produces that these particles lay on average closer to each other. Since as a function of the field intensity we have shown that the strength of the interaction is reduced, and that this happens no matter how weak the external field may be, we infer that a similar effect is taking place in QCD where due to asymptotic freedom, the strength of the interaction gets reduced as the virtual particles get closer to each other. This weakening of the interaction with proximity between the virtual particles that make up the vacuum should manifest itself as well in the weakening of the quark condensate with the field strength, as is also observed in lattice QCD around the critical temperature. We believe this description will play an important role in the interpretation of the lattice QCD results for the behavior of the critical temperature and the quark condensate with the field intensity as well as in determining the location of the CEP in QCD with and without magnetic fields.
%%%%%%%%%%%%%%%%%%%%%%%%%%%%%%
\begin{figure}[th]
\begin{center}
\includegraphics[scale=0.6]{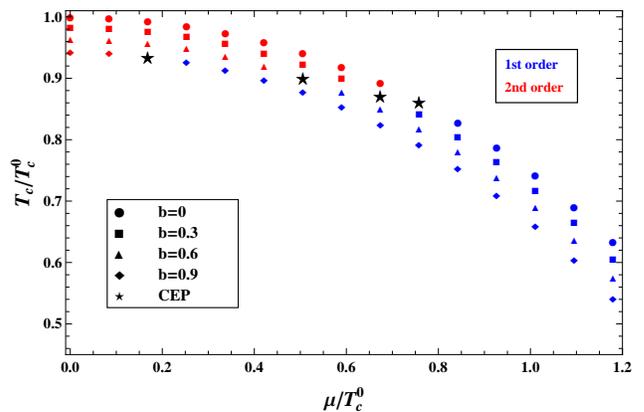}
\end{center}
\caption{Phase diagram in the temperature quark chemical potential plane computed with thermo-magnetic corrections to the couplings using the bare values $\lambda=0.36$, $g=0.51$ corresponding to $a/T_c^0=0.66$  for different values of the magnetic field strength. The phase transitions to the left (right) of the CEP in each case are of second (first) order.}
\label{fig8}
\end{figure}
%%%%%%%%%%%%%%%%%%%%%%%%%%%%%%

\section*{Acknowledgments}

This work has been supported in part by DGAPA-UNAM (Mexico) under grant number PAPIIT-IN101515, CONACyT (Mexico) under grant number 128534, FONDECYT (Chile) under grant  numbers 1130056 and 1120770, and 21110295, NRF (South Africa), and the Harry Oppenheimer Memorial Foundation (South Africa). R. Z. acknowledges support from CONICYT under Grant No. 21110295.

\section*{Appendix}

\subsection{Thermo-magnetic corrections to the boson coupling}

The thermo-magnetic correction to $\lambda$ involves the diagrams shown in Fig.~\ref{fig1}, column (a). It is only necessary to consider the case where the loop is made of charged pions (the bottom diagram in column (a) of Fig.~\ref{fig1}), since the other contributions can be obtained from this one after letting $B \rightarrow 0$. The calculation is carried out in the static limit, {\it i.e.} where $P_i=(\Pi,{\mbox{\bf{p}}}= 0)$. The explicit expression is given by  
\bea
J(P_i;m_i^2)&=&T\sum_n\int\frac{d^3k}{(2\pi)^3}D_B(P_i-K)D_B(K) \nn
&=&J_{n=0}(P_i;m_i^2) + J_{n\neq 0}(P_i;m_i^2).
\eea
First we consider the contribution from the zero mode
\bea
&&J_{n=0}(P_i;m_i^2)=T\int{\frac{d^3k}{(2\pi)^3}}\nn
&\times&\int_0^\infty ds \frac{e^{-s(\omega_n^2+(p_3-k_3)^2+(p_\perp-k_\perp)^2
   \frac{\tanh (qBs)}{qBs} + m^2)}}{\cosh (qBs)} \nn
&\times&\int_0^\infty d\tau \frac{e^{-\tau(k_3^2+k_\perp^2
   \frac{\tanh (qB\tau)}{qB\tau} + m^2)}}{\cosh (qB\tau)}.   
\eea
In the Hard Thermal Loop Approximation (HTL)  $P_3$ y $P_\perp$ are small quantities with respect to T, and the same occurs with the mass. In this way we find
\bea
&&J_{n=0}=T\int{\frac{d^3k}{(2\pi)^3}} \nn
&\times&\int_0^\infty ds \frac{e^{-s(\omega_n^2+k_3^2+k_\perp^2
   \frac{\tanh (qBs)}{qBs})}}{\cosh (qBs)} \nn
&\times&\int_0^\infty d\tau \frac{e^{-\tau(k_3^2+k_\perp^2
   \frac{\tanh (qB\tau)}{qB\tau})}}{\cosh (qB\tau)}.
\eea
Carrying out the integrals, we obtain
\bea
   J_{n=0}(\omega=\Pi)=\frac{T}{16\pi}\frac{1}{(2qB)^{1/2}}  \zeta \left(\frac{3}{2}, \frac{1}{2} + \frac{\Pi(T,\mu)}{2qB}\right).
\eea 
For the non-zero modes $(\omega _{n} \neq 0)$ we find
\bea
&&J_{n\neq 0}(P_i;m_i^2)=T\sum_{n \neq 0}\int{\frac{d^3k}{(2\pi)^3}} \nn
&&\biggl[ \biggl(\frac{1}{\omega_n^2+k^2+m^2} - \frac{(eB)^2}{(\omega_n^2+k^2+m^2)^3}  \nn
&& + \frac{2 (eB)^2 k_\perp^2 }{(\omega_n^2+k^2+m^2)^4}\biggr) \biggl(\frac{1}{(\omega-\omega_n)^2+(p-k)^2+m^2} \nn
&-& \frac{(eB)^2}{((\omega-\omega_n)^2+(p-k)^2+m^2)^3}  \nn
&& + \frac{2 (eB)^2 k_\perp^2 }{((\omega-\omega_n)^2+(p-k)^2+m^2)^4}\biggr)\biggr].
\eea
Still in the HTL approximation we find
\bea
&&J_{n\neq 0}(P_i;m_i^2)=T\sum_{n \neq 0}\int{\frac{d^3k}{(2\pi)^3}} \nn
&& \biggl[\biggl(\frac{1}{\omega_n^2+k^2} - \frac{(eB)^2}{2(\omega_n^2+k^2)^3} \biggr)  \nn
&& \biggl(\frac{1}{(\omega-\omega_n)^2+k^2} - \frac{(eB)^2}{((\omega-\omega_n)^2+k^2)^3}\biggr)\biggr].
\eea
Since we only consider terms up to order  ${\mathcal{O}}$ $(qB)^2$, we have
\bea
&&J_{n\neq 0}(P_i;m_i^2)=T\sum_{n \neq 0}\int{\frac{d^3k}{(2\pi)^3}} \nn
&& \biggl[\frac{1}{(\omega_n^2+k^2+\omega^2)^2} - \frac{(eB)^2}{(\omega_n^2+k^2+\omega^2)^4}   \biggr].
\eea
In order to calculate the above integrals we make use of dimensional regularization, and of the Mellin summation technique~\cite{Mellin}, to find
\bea
&&J_{n\neq 0}(\omega^2=\Pi)=-\frac{1}{16 \pi^2} \Big[\ln\left( \frac{(4\pi T)^2}{2a^2} \right) +1 - 2\gamma_E\nn
   &+&\zeta(3) \left(\frac{\sqrt{\Pi}}{2 \pi T}\right)^2   \Big]-\frac{(q B)^2 }{1024 \pi^6 T^4} \zeta(5).
\eea
Joining both contributions we find
\bea
   J(\omega=\Pi)&=&\frac{T}{16\pi}\frac{1}{(2qB)^{1/2}}  \zeta \left(\frac{3}{2}, \frac{1}{2} + \frac{\Pi}{2qB}\right) \nn
   &-&\frac{1}{16 \pi^2} \Big[\ln\left( \frac{(4\pi T)^2}{2a^2} \right) +1 - 2\gamma_E\nn
   &+&\zeta(3) \left(\frac{\sqrt{\Pi}}{2 \pi T}\right)^2   \Big]-\frac{(q B)^2 }{1024 \pi^6 T^4} \zeta(5).
   \label{Jfinal}
\eea 
In the case of the diagrams involving neutral bosons we have
\bea
I(P_i;m_i^2)&=&T\sum_n\int\frac{d^3k}{(2\pi)^3}D(P_i-K)D(K) \nn
&=&I_{n=0}(P_i;m_i^2) + I_{n\neq 0}(P_i;m_i^2).
\eea
In order to calculate $I(P_i;m_i^2)$,  we take limit $(qB) \rightarrow 0$ in Eq.~(\ref{Jfinal}). The limit of the Hurwitz Zeta function is not trivial and we we use the following assymptotic expansion~\cite{Paris}
\be
\zeta(s,y)=\frac{1}{2}y^{-s} + \frac{y^{1-2}}{s-1}+\sum_{k=1}^{\infty} \frac{B_{2k}}{(2k)!} \frac{\Gamma(2k+s-1)}{\Gamma(s)y^{2k+s-1}},
\label{expansionzeta}
\ee
where $B_{2k}$ are Bernoulli numbers. This expansion is valid for large values of $y$, which is equivalent to having a small value for $qB$. In our case $s=3/2$ and $y=\frac{1}{2} + \frac{\Pi}{2qB}$ and we find
\be
\zeta \left(\frac{3}{2}  , \frac{1}{2}+\frac{\Pi}{2qB}\right) \approx \frac{2 (2qB)^{1/2}}{\sqrt{\Pi}} - \frac{1}{16} \frac{(2qB)^{5/2}}{\Pi^{5/2}} + \cdots .
\ee
Using the above expansion in Eq.~(\ref{Jfinal}) we obtain
\bea
   I(\omega=\Pi)&=&\frac{T}{8\pi}\frac{1}{\sqrt{\Pi}}\nn
   &-&\frac{1}{16 \pi^2} \Big[\ln\left( \frac{(4\pi T)^2}{2a^2} \right) +1 - 2\gamma_E\nn
   &+&\zeta(3) \left(\frac{\sqrt{\Pi}}{2 \pi T}\right)^2   \Big].
\label{nneq02}
\eea

\subsection{Thermo-magnetic corrections to the fermion-boson coupling}

The determination of the thermo-magnetic correction to the coupling $g$ involves the diagram shown in Fig.~\ref{fig2} (a). We call the one-loop effective vertex $\Gamma$. The calculation is done up to order ${\mathcal{O}}$ $(qB)$
\bea
\Gamma&=&-g\nn
&+& g^3T\sum_{n}\int{\frac{d^3k}{(2\pi)^3}}\gamma_5 S(P_1-K) S(P_2-K)\gamma_5 D(K)\nn
&\equiv&-g(1+\delta\Gamma).
\eea
Concentrating on $\delta \Gamma$
\bea
\delta\Gamma &=&-g^2T\sum_{n}\int{\frac{d^3k}{(2\pi)^3}}\gamma_5 S(P_1-K) \nn
&\times&S(P_2-K)\gamma_5 D(K) \nn
&=&-g^2T\sum_{n}\int{\frac{d^3k}{(2\pi)^3}}\bigg[\frac{-(\slshh{P_1}-\slshh{K})}{(P_1-K)^2+m^2} \nn
&+& \frac{iqB\gamma_1\gamma_2(\gamma \cdot (P_2-K)_{||})}{[(P_2-K)^2+m^2]^2}  \biggr] \bigg[\frac{-(\slshh{P_2}-\slshh{K})}{(P_2-K)^2+m^2} \nn
&+& \frac{iqB\gamma_1\gamma_2(\gamma \cdot (P_2-K)_{||})}{[(P_2-K)^2+m^2]^2}  \biggr]\frac{1}{K^2+m_{\pi}^2}. 
\eea
In the HTL approximation, we get
\bea
\delta\Gamma &=&-g^2T\sum_{n}\int{\frac{d^3k}{(2\pi)^3}} \biggl[ (\slshh{K})^2\tilde{\Delta}(P_1-K) \tilde{\Delta}(P_2-K) \nn 
&\times&\Delta(K) -iqB \slshh{K}\gamma_1\gamma_2(\gamma \cdot K)_{||} \tilde{\Delta}(P_1-K)  \nn
&\times&\tilde{\Delta}^2(P_2-K)\Delta(K) - iqB \gamma_1\gamma_2(\gamma \cdot K)_{||} \slshh{K} \nn
&\times & \tilde{\Delta}(P_1-K)\tilde{\Delta}^2(P_2-K)\Delta(K) \biggr]\nn
&=& \delta \Gamma_{TV}+ \delta \Gamma_{TB},
\label{deltag}
\eea
where
\bea
\delta \Gamma_{TV}&=&-g^2T\sum_{n}\int{\frac{d^3k}{(2\pi)^3}}\slshh{K}\slshh{K}\nn
&\times& \tilde{\Delta}(P_1-K) \tilde{\Delta}(P_2-K) \Delta(K),
\eea
is the vacuum $+$ thermal contribution, and where
\bea
\delta\Gamma_{TB} &=&-g^2T\sum_{n}\int{\frac{d^3k}{(2\pi)^3}} \biggl[ -iqB \slshh{K}\gamma_1\gamma_2(\gamma \cdot K)_{||} \nn
&\times& \tilde{\Delta}(P_1-K)\tilde{\Delta}^2(P_2-K)\Delta(K) \nn
&-&iqB \gamma_1\gamma_2(\gamma \cdot K)_{||} \slshh{K} \tilde{\Delta}(P_1-K)\nn
&\times&\tilde{\Delta}^2(P_2-K)\Delta(K) \biggr],
\eea
is the thermo-magnetic contribution.  We now consider $\delta \Gamma_{TV}$ in the HTL approximation 
\bea
\delta \Gamma_{TV}&=&-g^2T\sum_{n}\int{\frac{d^3k}{(2\pi)^3}}\slshh{K}\slshh{K} \tilde{\Delta}(P_1-K) \nn
&\times& \tilde{\Delta}(P_2-K) \Delta(K) \nn
&&=g^2T\sum_{n}\int{\frac{d^3k}{(2\pi)^3}}\tilde{\Delta}(P_1-K)\tilde{\Delta}(P_2-K) \nn
&&=\frac{1}{8\pi^2}\left(\ln\left(\frac{a}{T\pi}\right)+\frac{\gamma_E}{2}-\frac{1}{2}-\ln(2\pi) \right).
\eea
Next, we concentrate on the last two terms of Eq. (\ref{deltag}), which make up the thermo-magnetic contribution  $\delta \Gamma_{TB}$. First, we recall 
\bea
\gamma_5=\gamma_4\gamma_1\gamma_2\gamma_3,
\eea
anti-commutes with the other gamma matrices. We introduce the decomposition  
\bea
   \gamma_1\gamma_2\slshh{K_{\|}}=\gamma_5\left[(K\cdot b)\slshh{u} - (K\cdot u)\slsh{b}\right],
\eea
where we have introduced the four-vectors
\bea
   u_\mu&=&(1,0,0,0)\nn
   b_\mu&=&(0,0,0,1).
\eea
We stress that in the HTL approximation, $P_1$ y $P_2$ are small quantities that can be considered of the same order. In this way the thermo-magnetic contribution can be written as 
\bea
&&\delta \Gamma_{TB}=-g^2T\sum_{n}\int{\frac{d^3k}{(2\pi)^3}}\tilde{\Delta}(P_1-K)\nn
&\times &\tilde{\Delta}^2(P_2-K)\Delta(K)[-2iqB\gamma_5(\slshh{u}(K\cdot b)) \nn
&-& \slsh{b} (K \cdot u )] \slshh{K}.
\eea
We define 
\bea
\tilde{G}(P_1,P_2)&=&T\sum_{n}\int{\frac{d^3k}{(2\pi)^3}}\tilde{\Delta}(P_1-K)\tilde{\Delta}^2(P_2-K)\nn
&\times &\Delta(K)[(\slshh{u}(K\cdot b)) - \slsh{b} (K \cdot u )] \slshh{K},
\eea
to obtain
\bea
\delta \Gamma_{TB} = 2ig^2(qB)\gamma_5 \tilde{G}(P_1,P_2),
\eea
where $\tilde{G}(P_1,P_2)$ can be expressed in terms of the  tensor ${\mathcal{J}}_{\alpha i}$ $(\alpha =1,\ldots 4,\ i=3,4)$ given by
\bea
   {\mathcal{J}}_{\alpha i}&=&T\sum_n\int\frac{d^3k}{(2\pi)^3}K_\alpha K_i\nn
   &\times&\widetilde{\Delta}^2(K)\Delta(P_1-K)\Delta(P_2-K).
   \label{mathcalJ}
\eea
In order to calculate the sum over Matsubara frequencies we use
\bea
   \widetilde{Y}_0&=&T\sum_n\widetilde{\Delta}^2(K)\Delta(P_1-K)\Delta(P_2-K)\nn
   &=&\left(-\frac{\partial}{\partial m^2}\right)T\sum_n\widetilde{\Delta}(K)\Delta(P_1-K)\Delta(P_2-K)\nn
   &\equiv&\left(-\frac{\partial}{\partial m^2}\right)\widetilde{X}_0\nn
   \widetilde{Y}_1&=&T\sum_n\omega_n\widetilde{\Delta}^2(K)\Delta(P_1-K)\Delta(P_2-K)\nn
   &=&\left(-\frac{\partial}{\partial m^2}\right)T\sum_n
   \omega_n\widetilde{\Delta}(K)\Delta(P_1-K)\Delta(P_2-K)\nn
   &\equiv&\left(-\frac{\partial}{\partial m^2}\right)\widetilde{X}_1\nn
   \widetilde{Y}_2&=&T\sum_n\omega_n^2\widetilde{\Delta}^2(K)\Delta(P_1-K)\Delta(P_2-K)\nn
   &=&\left(-\frac{\partial}{\partial m^2}\right)T\sum_n\omega_n^2\widetilde{\Delta}(K)\Delta(P_1-K)\Delta(P_2-K)\nn
   &\equiv&\left(-\frac{\partial}{\partial m^2}\right)\widetilde{X}_2,
\label{sums}
\eea
where $\widetilde{X}_0$, $\widetilde{X}_1$, $\widetilde{X}_2$ are given by
\bea
   \widetilde{X}_0&=&-\sum_{s,s_1,s_2}\frac{ss_1s_2}{8EE_1E_2}
   \frac{1}{i(\omega_1-\omega_2)-s_1E_1+s_2E_2}\nn
   &\times&\left[\frac{1-\widetilde{f}(sE)+f(s_1E_1)}{i\omega_1-sE-s_1E_1} - 
   \frac{1-\widetilde{f}(sE)+f(s_2E_2)}{i\omega_2-sE-s_2E_2}
      \right]\nn
    \widetilde{X}_1&=&i\sum_{s,s_1,s_2}\frac{s_1s_2E}{8EE_1E_2}
   \frac{1}{i(\omega_1-\omega_2)-s_1E_1+s_2E_2}\nn
   &\times&\left[\frac{1-\widetilde{f}(sE)+f(s_1E_1)}{i\omega_1-sE-s_1E_1} - 
   \frac{1-\widetilde{f}(sE)+f(s_2E_2)}{i\omega_2-sE-s_2E_2}
       \right]\nn
   \widetilde{X}_2&=&\sum_{s,s_1,s_2}\frac{ss_1s_2E^2}{8EE_1E_2}
   \frac{1}{i(\omega_1-\omega_2)-s_1E_1+s_2E_2}\nn
   &\times&\left[\frac{1-\widetilde{f}(sE)+f(s_1E_1)}{i\omega_1-sE-s_1E_1} - 
   \frac{1-\widetilde{f}(sE)+f(s_2E_2)}{i\omega_2-sE-s_2E_2}
       \right].\nn
\label{tildeXs}
\eea
The leading temperature behavior is obtained from the terms with $s=-s_1=-s_2$. We consider in detail the calculation of $\widetilde{X}_0$ for those terms and make the approximation where $f(E_1)\simeq f(E_2)\simeq f(E)$, namely, that the Bose-Einstein distribution depends on $E=\sqrt{k^2+m^2}$ and thus on the quark mass. This approximation allows to find the leading temperature behavior for $m\rightarrow 0$, since it amounts to keep the quark mass as an infrared regulator. Also, using that $E_i\simeq k - \vec{p}_i\cdot\hat{k}$, $i=1,2$, we find  
\bea
   \widetilde{X}_0&\simeq&-\frac{1}{8k^2}\frac{\left[\widetilde{f}(E)+f(E)\right]}{E}\nn
   &\times&\left\{
   \frac{1}{(i\omega_1 + \vec{p}_1\cdot\hat{k})(i\omega_2 + \vec{p}_2\cdot\hat{k})} \right.\nn
   &+& \left.
   \frac{1}{(i\omega_1 - \vec{p}_1\cdot\hat{k})(i\omega_2 - \vec{p}_2\cdot\hat{k})}
   \right\},
\label{HTLX0}
\eea
where we have set $E_1=E_2=k$ in the denominator of the first fraction. Similarly
\bea
   \widetilde{X}_1&\simeq&-\frac{i}{8k}\frac{\left[\widetilde{f}(E)+f(E)\right]}{E}\nn
   &\times&\left\{
   \frac{1}{(i\omega_1 + \vec{p}_1\cdot\hat{k})(i\omega_2 + \vec{p}_2\cdot\hat{k})}\right. \nn
   &-& \left.
   \frac{1}{(i\omega_1 - \vec{p}_1\cdot\hat{k})(i\omega_2 - \vec{p}_2\cdot\hat{k})}
   \right\}\nn
   \widetilde{X}_2&\simeq&\frac{1}{8}\frac{\left[\widetilde{f}(E)+f(E)\right]}{E}\nn
   &\times&\left\{
   \frac{1}{(i\omega_1 + \vec{p}_1\cdot\hat{k})(i\omega_2 + \vec{p}_2\cdot\hat{k})} \right.\nn
   &+& \left.
   \frac{1}{(i\omega_1 - \vec{p}_1\cdot\hat{k})(i\omega_2 - \vec{p}_2\cdot\hat{k})}
   \right\}.
\label{HTLX1}
\eea
Using Eqs.~(\ref{HTLX0}) and~(\ref{HTLX1}) in Eqs.~(\ref{sums}) and~(\ref{mathcalJ}), we find
\bea
   {\mathcal{J}}_{\alpha i}&=&-\frac{1}{8\pi^2}\left(-\frac{\partial}{\partial y^2}\right)\int_0^\infty 
   \frac{dx\ x^2}{\sqrt{x^2+y^2}}\nn
   &\times&\left[\widetilde{f}(\sqrt{x^2+y^2})+f(\sqrt{x^2+y^2})\right]\nn
   &\times&\int\frac{d\Omega}{4\pi}
   \frac{\hat{K}_\alpha \hat{K}_i}{(P_1\cdot\hat{K})(P_2\cdot\hat{K})},
\label{integr}
\eea
where we have defined $x=k/T$, $y=m/T$, $\hat{K}=(-i,\hat{k})$, $P_1=(-\omega_1,\vec{p}_1)$ and $P_2=(-\omega_2,\vec{p}_2)$. The integrals over $x$ can be expressed in terms of the well known functions~\cite{Jackiw}
\bea
   h_n(y)&=&\frac{1}{\Gamma (n)}\int_0^\infty 
   \frac{dx\ x^{n-1}}{\sqrt{x^2+y^2}}\frac{1}{e^{\sqrt{x^2+y^2}}-1}\nn
   f_n(y)&=&\frac{1}{\Gamma (n)}\int_0^\infty 
   \frac{dx\ x^{n-1}}{\sqrt{x^2+y^2}}\frac{1}{e^{\sqrt{x^2+y^2}}+1},
\label{handf}
\eea
which satisfy the differential equations
\bea
   \frac{\partial h_{n+1}}{\partial y^2}&=&-\frac{h_{n-1}}{2n}\nn
   \frac{\partial f_{n+1}}{\partial y^2}&=&-\frac{f_{n-1}}{2n},
\label{diffeq}
\eea
therefore
\bea
   {\mathcal{J}}_{\alpha i}&=&-\frac{1}{16\pi^2}\left[h_1(y)+f_1(y)\right]\nn
   &\times&\int\frac{d\Omega}{4\pi}
   \frac{\hat{K}_\alpha \hat{K}_i}{(P_1\cdot\hat{K})(P_2\cdot\hat{K})}.
\label{simpl}
\eea
Using the high temperature expansions for $h_1(y)$ and $f_1(y)$~\cite{Kapusta}
\bea
   h_1(y)&=&\frac{\pi}{2y} + \frac{1}{2}\ln\left(\frac{y}{4\pi}\right) + \frac{1}{2}\gamma_E + \ldots\nn
   f_1(y)&=&-\frac{1}{2}\ln\left(\frac{y}{\pi}\right) - \frac{1}{2}\gamma_E + \ldots,
\label{expansions}
\eea
and keeping the leading terms, we obtain
\bea
   {\mathcal{J}}_{\alpha i}=\frac{1}{16\pi^2}\left[\ln(2) - \frac{\pi}{2}\frac{T}{\sqrt{\Pi}}\right]
   \int\frac{d\Omega}{4\pi}
   \frac{\hat{K}_\alpha \hat{K}_i}{(P_1\cdot\hat{K})(P_2\cdot\hat{K})}.
\label{finally}
\eea
Hence
\bea
&&\delta \Gamma_{TB}=-\frac{2ig^2(qB)\gamma_5}{16\pi^2} \left[ \ln(2)- \frac{\pi}{2}\frac{T}{\sqrt{\Pi}} \right] \nn
&\times& \int{\frac{d\Omega}{4\pi}} \frac{[(\slshh{u}(\hat{K}\cdot b)) - \slsh{b} (\hat{K} \cdot u )] \slshh{\hat{K}}}{(P_1 \cdot \hat{K})(P_2 \cdot \hat{K})}.
\label{Gmuexpl}
\eea
In order to consider the thermo-magnetic dependence of the fermion-boson coupling we consider explicitly the quantities appering on the r.h.s of Eq.~(\ref{Gmuexpl})
\bea
   J_{\alpha i}(P_1,P_2)\equiv\int\frac{d\Omega}{4\pi}\frac{\hat{K}_\alpha\hat{K}_i}
   {(P_1\cdot{\hat{K}})(P_2\cdot{\hat{K}})},
\label{Js}
\eea
For simplicity we choose a configuration where the momenta $\vec{p}_1$ and $\vec{p}_2$ form a relative angle $\theta_{12}=\pi$. This configuration corresponds, for instance, to a thermal gluon decaying into a quark-antiquark pair in the center of mass system, and is therefore general enough. Consider first $J_{44}(P_1,P_2)$
\bea
   J_{44}(P_1,P_2)&=&-\frac{1}{2}\frac{1}{i\omega_1p_2+i\omega_2p_1}\nn
   &\times&
   \int_{-1}^1dx\left\{\frac{p_1}{i\omega_1 + p_1x} + \frac{p_2}{i\omega_2 - p_2x}\right\}\nn
   &=&-\frac{1}{2}\frac{1}{i\omega_1p_2+i\omega_2p_1}\nn
   &\times&
   \left\{ \ln\left(\frac{i\omega_1 + p_1}{i\omega_1 - p_1}\right) + \ln\left(\frac{i\omega_2 + p_2}{i\omega_2 - p_2}\right)
   \right\}.\nn
\label{J44}
\eea
We now perform the analytic continuation to Minkowski space $i\omega_{1,2}\rightarrow p_{0 1,0 2}$ [$\hat{K}\rightarrow (-1,\hat{k})$], and consider the scenario where $p_{0 1}=p_{0 2}\equiv p_0$ and $p_1=p_2\equiv p$, leading to
\bea
   J_{44}\rightarrow J_{00}=\frac{1}{2p_0p}\ln\left(\frac{p_0+p}{p_0-p}\right).
\label{J44toMink}
\eea
Furthermore, we consider the {\it static limit} where the quarks are almost at rest, namely $p\rightarrow 0$, to find 
\bea
   J_{00}\stackrel{p\rightarrow 0}{\longrightarrow}\frac{1}{p_0^2}.
\label{J00inthelim}
\eea
Now we consider $J_{33}(P_1,P_2)$ in the same momenta  configuration
\bea
   J_{33}(P_1,P_2)&=&\frac{1}{2}\frac{1}{i\omega_1p_2+i\omega_2p_1}\nn
   &\times&
   \int_{-1}^1dx\ x^2\left\{\frac{p_1}{i\omega_1 + p_1x} + \frac{p_2}{i\omega_2 - p_2x}\right\}\nn
   &=&-\frac{1}{i\omega_1p_2+i\omega_2p_1}\nn
   &\times& \left\{
   \frac{i\omega_1}{p_1}\left[ 1 - \frac{i\omega_1}{2p_1}\ln\left(\frac{i\omega_1 + p_1}{i\omega_1 - p_1}\right)\right]
   \right.\nn
   &+& \left. \frac{i\omega_2}{p_2}\left[ 1 - \frac{i\omega_2}{2p_2}\ln\left(\frac{i\omega_2 + p_2}{i\omega_2 - p_2}\right)\right]
   \right\}.\nn
\label{J33inic}
\eea
After analytical continuation to Minkowski space and in the same scenario where $p_{0 1}=p_{0 2}\equiv p_0$ and $p_1=p_2\equiv p$, we obtain
\bea
   J_{33}=-\frac{1}{p^2}\left[1 - \frac{p_0}{2p}\ln\left(\frac{p_0+p}{p_0-p}\right)\right].
\label{J33after}
\eea
In the limit where $p\rightarrow 0$ this gives
\bea
    J_{33}\stackrel{p\rightarrow 0}{\longrightarrow}\frac{1}{3p_0^2}.
\label{J33}
\eea
In this same limit, $p \rightarrow 0$, we find
\bea
\delta\Gamma_{TB}&=&2g^2(J_{33}+J_{44})q \vec{\Sigma} \cdot \vec{B}\frac{\left[ \ln(2)- \frac{\pi}{2}\frac{T}{\sqrt{\Pi}} \right] }{16\pi^2} \nn
&=&2g^2\left(\frac{4}{3 p_0^2}\right)\frac{\left[ \ln(2)- \frac{\pi}{2}\frac{T}{\sqrt{\Pi}} \right] }{16\pi^2}q \vec{\Sigma} \cdot \vec{B},
\eea
where
\bea
\vec{\Sigma} \cdot \vec{B}=i\gamma_1\gamma_2B.
\eea
By taking both contributions into account, $\delta \Gamma_{TV}+ \delta \Gamma_{TB}$, we can find $\Gamma$. Considering the contributions of the \textit{up-} and \textit{down-}quarks explicitly, and taking $p_0^2 \rightarrow m_f^2$ we finally obtain the correction to the coupling $g$.
\bea
g_{\mbox{\small{eff}}}=g(1+g^2(g_{\mbox{\small{TB}}}+g_{\mbox{\small{TV}}})),
\eea
where
\bea
g_{\mbox{\small{TB}}}&=&\frac{(q_u+q_d)B}{8\pi^2}\left(\frac{4}{3mf^2}\right)\left(\ln(2)-\frac{\pi T}{2\sqrt{\Pi}}\right)\nn
g_{\mbox{\small{TV}}}&=&\frac{1}{8\pi^2}\left(\ln\left(\frac{a}{T\pi}\right)+\frac{\gamma_E}{2}-\frac{1}{2}-\ln(2\pi )   \right).
\eea

\subsection{Fermion thermal and density dependent mass}

\begin{figure}
	\centering
		\includegraphics[scale=0.45]{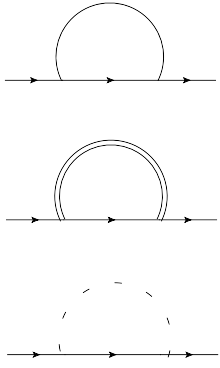}
		\caption{Diagrams contributing to the calculation of the fermion thermal mass.}
	\label{fig9}
\end{figure}

This calculation involves the three diagrams shown in Fig.~\ref{fig9}. We only consider the first one, since the computation of the other two diagrams is completely equivalent. We call this diagram $\Sigma_{\sigma}$
%%%%%%%%%%%%%%%%%%%%%%%%%
%%%%%%%%%%%%%%%%%%%%
\bea
\Sigma_{\sigma} = -g^2 \int{\frac{d^4K}{(2\pi)^4}} S(P-K)\Delta(K).
\eea
In the HTL approximation this gives
\bea
\Sigma_{\sigma} = -g^2 \int{\frac{d^4K}{(2\pi)^4}} \slshh{K}\Delta(K)\tilde{\Delta}(P-K),
\eea
where
\bea
\Delta(Q)&=&\frac{1}{\omega_n+k^2} \nn
\tilde{\Delta}(Q)&=&\frac{1}{\tilde{\omega}_n+k^2}.
\eea
The determination of $\Sigma_{\sigma}$ involves two kinds of Matsubara sums
\bea
\Sigma_{\sigma_{i=1,2,3}}&=&T\sum_n \Delta(K)\tilde{\Delta}(P-K)  \nn
\Sigma_{\sigma_{i=4}}&=& T\sum_n \omega_n \Delta(K)\tilde{\Delta}(P-K).
\eea
The first case refers to  $i=1,2,3$ and the second case to $i=4$. The first kind of contribution is
\bea
\Sigma_{\sigma_{i=1,2,3}}&=&T\sum_n \int{\frac{d^3K}{(2\pi)^3}} K_i\Delta(K)\tilde{\Delta}(P-K)=\nn
&-&\frac{1}{2\pi^2}\int{\frac{d\Omega}{4\pi}}\hat{K_i} \int{dk}K\biggl[\frac{f(k)+\tilde{f}(k-\mu)}{i\omega+\hat{k} \cdot \vec{p}} \nn
&-& \frac{f(k)+\tilde{f}(k+\mu)}{i\omega-\hat{k} \cdot \vec{p}} \biggr].
\eea
We have to deal with both the radial as well as the angular part. For the radial contribution we find
\bea
\int{dK} \ K [f(k)+\tilde{f}(k\pm \mu)] = \frac{\pi^2 T^2}{6} - T^2 Li_{2}(-e^{\mp \mu/T}). \nn
\eea
For the angular part we notice that the integral is symmetric under the transformation 
\bea
\hat{k} \rightarrow -\hat{k} ; d\Omega \rightarrow d\Omega,
\eea
implying 
\bea
\int{\frac{d\Omega}{4\pi}}\frac{1}{i\tilde{\omega}-\hat{k} \cdot \vec{p}} \rightarrow - \int{\frac{d\Omega}{4\pi}}\frac{1}{i\tilde{\omega}+\hat{k} \cdot \vec{p}}.
\eea
Therefore, we finally find
\bea
\Sigma_{\sigma_{i=1,2,3}}&=&-\frac{1}{8\pi^2} \biggl[2\left(\frac{\pi^2T^2}{6}\right) -T^2Li_2(-e^{\mu/T}) \nn
&-& T^2Li_2(-e^{-\mu/T}) \biggr] \int{\frac{d\Omega}{4\pi}}\frac{1}{i\tilde{\omega}+\hat{k} \cdot \vec{p}}.
\eea
The second contribution  $\Sigma_{\sigma_{i=4}}$ is
\bea
\Sigma_{\sigma_{i=4}}&=&T\sum_n \int{\frac{d^3K}{(2\pi)^3}} \omega_n\Delta(K)\tilde{\Delta}(P-K)=\nn
&-&\frac{i}{8\pi^2}\int{\frac{d\Omega}{4\pi}}\int{dk}K\biggl[\frac{f(k)+\tilde{f}(k-\mu)}{i\omega+\hat{k} \cdot \vec{p}} \nn
&+& \frac{f(k)+\tilde{f}(k+\mu)}{i\omega-\hat{k} \cdot \vec{p}} \biggr].
\eea
Carrying out the radial and the angular integrals, in a completely analogous way as before, we have 
\bea
\Sigma_{\sigma_{i=4}}&=&T\sum_n \int{\frac{d^3K}{(2\pi)^3}} \omega_n\Delta(K)\tilde{\Delta}(P-K)\nn
&=&\frac{-1}{8\pi^2}\biggl[2\left(\frac{\pi^2T^2}{6}\right) -T^2Li_2(-e^{\mu/T}) \nn
&-& T^2Li_2(-e^{-\mu/T}) \biggr]\int{\frac{d\Omega}{4\pi}}\frac{1}{i\tilde{\omega}+\hat{k} \cdot \vec{p}}.
\eea
The sum of the contribution of the different diagrams finally yields the thermal and density dependent correction to the fermion mass
\bea
(m_f^{\mbox{\tiny{them}}})^2=g^2 T^2\left(\frac{1}{3}-\frac{Li_2(-e^{\mu/T})}{\pi^2}-\frac{Li_2(-e^{-\mu/T})}{\pi^2}\right).\nn
\eea

\end{document}